\documentclass[11pt]{article}
\usepackage{amsmath,latexsym,amssymb,epsfig,psfrag,subfigure,graphicx,verbatim,multirow,appendix}
\makeatletter
\long\def\@caption#1[#2]#3{\par\addcontentsline{\csname
  ext@#1\endcsname}{#1}{\protect\numberline{\csname
  the#1\endcsname}{\ignorespaces #2}}\begingroup
    \small
    \@parboxrestore
    \@makecaption{\csname fnum@#1\endcsname}{\ignorespaces #3}\par
  \endgroup}
\makeatother
\newcommand{\newc}{\newcommand}
\newc{\lat}{{\ell at}}
\newc{\one}{{\bf 1}}
\newc{\mgut}{M_{\rm GUT}}
\newc{\mzero}{m_0}
\newc{\mhalf}{M_{1/2}}
\newc{\five}{{\bf 5}}
\newc{\fivebar}{{\bf\bar 5}}
\newc{\ten}{{\bf 10}}
\newc{\tenbar}{{\bf\bar{10}}}
\newc{\sixteen}{{\bf 16}}
\newc{\sixteenbar}{{\bf\bar{16}}}
\newc{\gsim}{\lower.7ex\hbox{$\;\stackrel{\textstyle>}{\sim}\;$}}
\newc{\lsim}{\lower.7ex\hbox{$\;\stackrel{\textstyle<}{\sim}\;$}}
\newc{\gev}{\,{\rm GeV}}
\newc{\mev}{\,{\rm MeV}}
\newc{\ev}{\,{\rm eV}}
\newc{\kev}{\,{\rm keV}}
\newc{\tev}{\,{\rm TeV}}
\newc{\mz}{m_Z}
\newc{\mw}{m_W}
\newc{\mpl}{M_{Pl}}
\newc{\mh}{m_h}
\newc{\mA}{m_A}
\newc{\tr}{\mbox{Tr}}

\newc\CO{\order}
\newc\CL{{\cal L}}
\newc\CY{{\cal Y}}
\newc\CH{{\cal H}}
\newc\CM{{\cal M}}
\newc\CF{{\cal F}}
\newc\CD{{\cal D}}
\newc\CN{{\cal N}}
\newc{\eps}{\epsilon}
\newc{\re}{\mbox{Re}\,}
\newc{\im}{\mbox{Im}\,}
\newc{\invpb}{\,\mbox{pb}^{-1}}
\newc{\invfb}{\,\mbox{fb}^{-1}}
\newc{\yddiag}{{\bf D}}
\newc{\yddiagd}{{\bf D^\dagger}}
\newc{\yudiag}{{\bf U}}
\newc{\yudiagd}{{\bf U^\dagger}}
\newc{\yd}{{\bf Y_D}}
\newc{\ydd}{{\bf Y_D^\dagger}}
\newc{\yu}{{\bf Y_U}}
\newc{\yud}{{\bf Y_U^\dagger}}
\newc{\ckm}{{\bf V}}
\newc{\ckmd}{{\bf V^\dagger}}
\newc{\ckmz}{{\bf V^0}}
\newc{\ckmzd}{{\bf V^{0\dagger}}}
\newc{\X}{{\bf X}}
\newc{\bbbar}{B^0-\bar B^0}

\newc{\sgn}{\mbox{sgn}\,}
\newc{\m}{{\bf m}}
\newc{\msusy}{M_{\rm SUSY}}
\newc{\munif}{M_{\rm unif}}
\newc{\slepton}{{\tilde\ell}}
\newc{\Slepton}{{\tilde L}}
\newc{\sneutrino}{{\tilde\nu}}
\newc{\selectron}{{\tilde e}}
\newc{\stau}{{\tilde\tau}}
\newc{\vckm}{V_{\!\mbox{\tiny CKM}}}
\def\bar#1{\overline{#1}}

\def\beq{\begin{equation}}
\def\eeq{\end{equation}}
\def\bea{\begin{eqnarray}}
\def\eea{\end{eqnarray}}
\newc{\ie}{{\it i.e.}}          \newc{\etal}{{\it et al.}}
\newc{\eg}{{\it e.g.}}          \newc{\etc}{{\it etc.}}
\newc{\cf}{{\it c.f.}}
\def\Dsl{\,\raise.15ex\hbox{/}\mkern-13.5mu D} 
\def\delsl{\raise.15ex\hbox{/}\kern-.57em\partial}
\def\Ksl{\hbox{/\kern-.6000em\rm K}}
\def\Asl{\hbox{/\kern-.6500em \rm A}}
\def\Dsl{\hbox{/\kern-.6000em\rm D}} 
\def\Qsl{\hbox{/\kern-.6000em\rm Q}}
\def\gradsl{\hbox{/\kern-.6500em$\nabla$}}
\begin{document}
\begin{titlepage}
\begin{flushright}
IFT-UAM/CSIC-11-61
\end{flushright}
\begin{center} ~~\\
\vspace{0.5cm} 
\Large {\bf\Large Upper bounds on SUSY masses from the LHC} 
\vspace*{1.5cm}

\normalsize{
{\bf  M. E. Cabrera$^a$\footnote[1]{maria.cabrera@uam.es},
J.A. Casas$^{a,b}$\footnote[2]{alberto.casas@uam.es}
and A. Delgado$^b$\footnote[3]{antonio.delgado@nd.edu} } \\

\smallskip  \medskip
$^a$\emph{Instituto de F\'\i sica Te\'orica, IFT-UAM/CSIC,}\\
\emph{U.A.M., Cantoblanco, 28049 Madrid, Spain}\\
$^b${\it Department of Physics, University of Notre Dame,}\\
{\it Notre Dame, IN 46556, USA}}

\medskip

\vskip0.6in 

\end{center}

\centerline{\large\bf Abstract}
\vspace{.5cm}
\noindent

The LHC is already putting bounds on the Higgs mass. In this paper we use those bounds to put constrains on the MSSM parameter space coming from the fact that, in supersymmetry, the Higgs mass is a function of the masses of sparticles, and therefore an upper bound on the Higgs mass translates into an upper bound for the masses for superpartners. We show that, although current bounds do not constrain yet the MSSM parameter space from above, once the Higgs mass bound improves big regions of this parameter space will be excluded, putting upper bounds on SUSY masses. On the other hand, for the case of split-SUSY we show that, for moderate or large $\tan\beta$, the present bounds on the Higgs mass already imply that the common mass for scalars cannot be greater than $10^{11}$ GeV. We show how these bounds will evolve as LHC continues to improve the limits on the Higgs mass.

\vspace*{2mm}
\end{titlepage}


The experimental bounds on the Higgs mass are rapidly changing. Besides the LEP lower bound, $m_h > 114.4$ GeV \cite{LEPHiggs}, LHC has recently extended the 95\% CL {\em excluded} region around 2$M_W$ to $149\ {\rm GeV}<m_h<190\ {\rm GeV}$, and has excluded a new range at $295\ {\rm GeV}<m_h<450\ {\rm GeV}$ \cite{LHC}. For sure we are likely to see stronger limits (and hopefully a discovery) as the LHC luminosity keeps growing.

These bounds put constrains on the parameter space of the Standard Model: they directly translate into bounds on the self-coupling of the Higgs. At tree-level the relation reads $m_h^2=2\lambda_{\rm tree} v^2$, where $\lambda$ is the SM Higgs quartic-self-coupling and $v=174.1$ GeV is the Higgs expectation value. This is not especially challenging per-se, in the sense that there is no particular prediction for that coupling in the pure SM. On the other hand, there are models where $\lambda$ is not a free parameter, but it is related to other parameters of the theory; hence a bound on $m_h$ can put constrains on those parameters. One classic example is supersymmetry (SUSY) where, in the minimal model (MSSM), $\lambda_{\rm tree}=\frac{1}{4}(g^2 +g'^2)\cos^2 2\beta$. Here $g, g'$ are the $SU(2)\times U(1)$ gauge couplings and $\tan \beta=\langle H_u\rangle/\langle H_d\rangle$, i.e. the ratio of expectation values of the two MSSM Higgs fields. This relation means, in particular, that at tree-level the mass of the Higgs in the MSSM is bounded by the mass of the $Z$-boson ($91.1$ GeV). As it is well-known, radiative corrections increase $m_h$, which can then get compatible with its experimental (LEP) lower bound, at the expense of requiring a relatively heavy spectrum ($\gsim 1$ TeV) of superpartners, which in turn introduces some fine tuning. Much work has been devoted to this important feature of the MSSM \cite{lista, Casas, Mariano, Haber}.

The approach of this Letter is \emph{opposite} and complementary: using the \emph{upper} bound on the Higgs mass to put an upper bound on the masses of supersymmetric particles.

It is common to hear that ``SUSY cannot be ruled out", meaning that one can always increase the masses of superpartners to avoid its discovery at the LHC or any conceivable experiment. However, for the above-mentioned reasons, in the MSSM the Higgs mass cannot be arbitrarily large. Actually, the radiative correction to $m_h^2$ depends logarithmically on the SUSY masses (principally on stop masses). This is easy to understand. The MSSM tree-level relation, $\lambda_{\rm tree}=\frac{1}{4}(g^2 +g'^2)\cos^2 2\beta$, breaks down at the threshold scale where supersymmetric particles become decoupled. Below that SUSY-threshold, $\lambda$ runs down to the electroweak scale following the SM renormalization group (RG) equation. The SUSY-threshold scale is essentially given by (an average of the two) stop masses, since, in the one-loop effective potential they are responsible for the largest contribution to the Higgs quartic-coupling to be matched with the SM-effective-theory (for details see e.g. \cite{Casas}). Hence, the enhancement of $\lambda$, and thus of $m_h^2$, goes logarithmically with the ratio of the SUSY-threshold scale to the electroweak scale.

Since the value of $\lambda$ at the SUSY-threshold scale is always perturbative, the Higgs mass in the MSSM necessarily obeys the SM perturbativity upper bound. For the extreme case when the supersymmetric masses are as large as $M_P$, and so is the threshold of new physics, this bound reads $m_h\lsim 180$ GeV \cite{upperSM}, which is already overtaken by the recent LHC exclusion ranges on $m_h$ quoted above. In other words, for the MSSM the only relevant experimentally allowed range for $m_h$ is
\bea
114.4\ {\rm GeV}<m_h<149\ {\rm GeV}\ .
\eea
These upper and lower bounds on $m_h$ translate, respectively, into upper and lower bounds on the masses of the supersymmetric particles. The latter can be complemented with the recent direct LHC bounds on the size of the supersymmetric masses \cite{LHCdirect}, giving the window of scales where the MSSM can live. In this letter we are going to show explicitly this window, describing how it will evolve as the LHC improves the limits on the mass of the Higgs or discovers its existence. We will see that the upper bounds on the MSSM scale are not yet of physical significance, but they will get much stronger in the near future. 

In this paper we have evaluated the Higgs mass, starting with the tree-level value of $\lambda$ at SUSY-threshold scale, corrected with 1-loop threshold corrections
\bea
\lambda (M_{\rm SUSY})= \frac{1}{4}(g^2 + g'^2)\cos^2 2\beta  + \frac{3}{16\pi^2}\frac{m_t^4}{v^4}X_t
\label{lambda}
\eea

\noindent
with
 \bea
 X_t=\frac{2 (A_t-\mu\cot\beta)^2}{M^2_{\rm SUSY}} \left (1-
 \frac{(A_t-\mu\cot\beta)^2}{12 M^2_{\rm SUSY}}
 \right)
 \eea
 
 \noindent 
in where $m_t$ is the running top mass corresponding to a pole mass $M_t=173.1\pm 1.25$ GeV \cite{Mt}; $A_t$ is the trilinear scalar coupling, $\mu$ is the mass parameter for the Higgses in the superpotential; and $M_{\rm SUSY}$ represents a certain average of the stop masses \cite{Casas, Mariano, Haber}. Note that, unless the combination $\left|A_t-\mu\cot\beta\right|$ becomes larger than $\sqrt{12}M_{\rm SUSY}$, which is very odd (and almost in coflict with charge and color breaking bounds), the value of $X_t$ is in the range $0\leq X_t\leq 6$.

The above value of $\lambda$ has to be run down to the electroweak scale, say $Q_{\rm ew}$. Then the Higgs mass is given by
\bea
m_h \simeq 2 \lambda(Q_{\rm ew})v^2
\eea
This relation gets one-loop radiative corrections (in the effective SM theory), which are rendered negligible by choosing appropriately $Q_{\rm ew}$. A nearly optimal choice is $Q_{\rm ew}=M_t$ \cite{Casas}. Finally, to get the pole mass, $M_h$, one has to add (pretty small) radiative corrections. We have performed the previously-sketched calculation of $m_h$ using the 2-loop SM RG equation of $\lambda$, which is coupled to the RG equations of the other SM parameters, in especial the top Yukawa coupling, $y_t$, and the strong coupling, $\alpha_3$. The complete set of RG equations can be found e.g. in ref.\cite{RGs}.

Let us mention that in the literature there are approximate analytic formulae for the Higgs mass, obtained by approximating the running of $\lambda$ at a certain order in the leading log expansion \cite{Mariano, Haber}. The first terms of those formulae read 
\bea
\label{mhaprox}
m_h^2\simeq M_Z^2\cos^2 2\beta + \frac{3}{4\pi^2}\frac{m_t^4}{v^2}\left[\frac{1}{2} X_t+\log\frac{M^2_{\rm SUSY} }{M^2_t}\right]+\cdots
\label{mass}
\eea
These approximations are only valid for $M_{\rm SUSY}\lsim 1$ TeV, so they are {\em not} applicable to our problem. However, eq.(\ref{mhaprox}) is useful to qualitatively understand the numerical results, so we have shown it explicitly.

\begin{figure}[t]
\begin{center}
\includegraphics[width=0.45\linewidth]{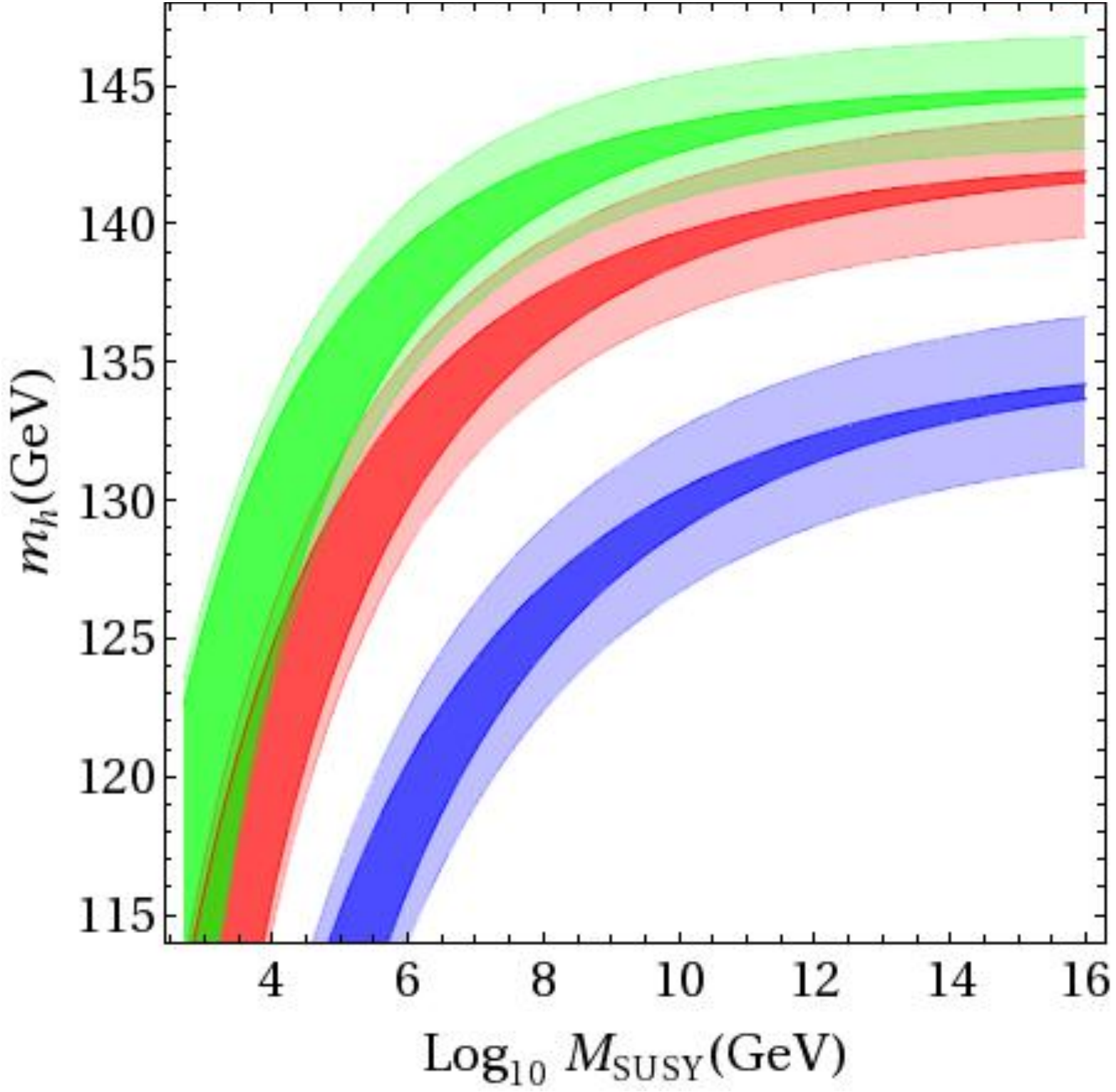} %
\hspace{0.05\linewidth} %
\includegraphics[width=0.45\linewidth]{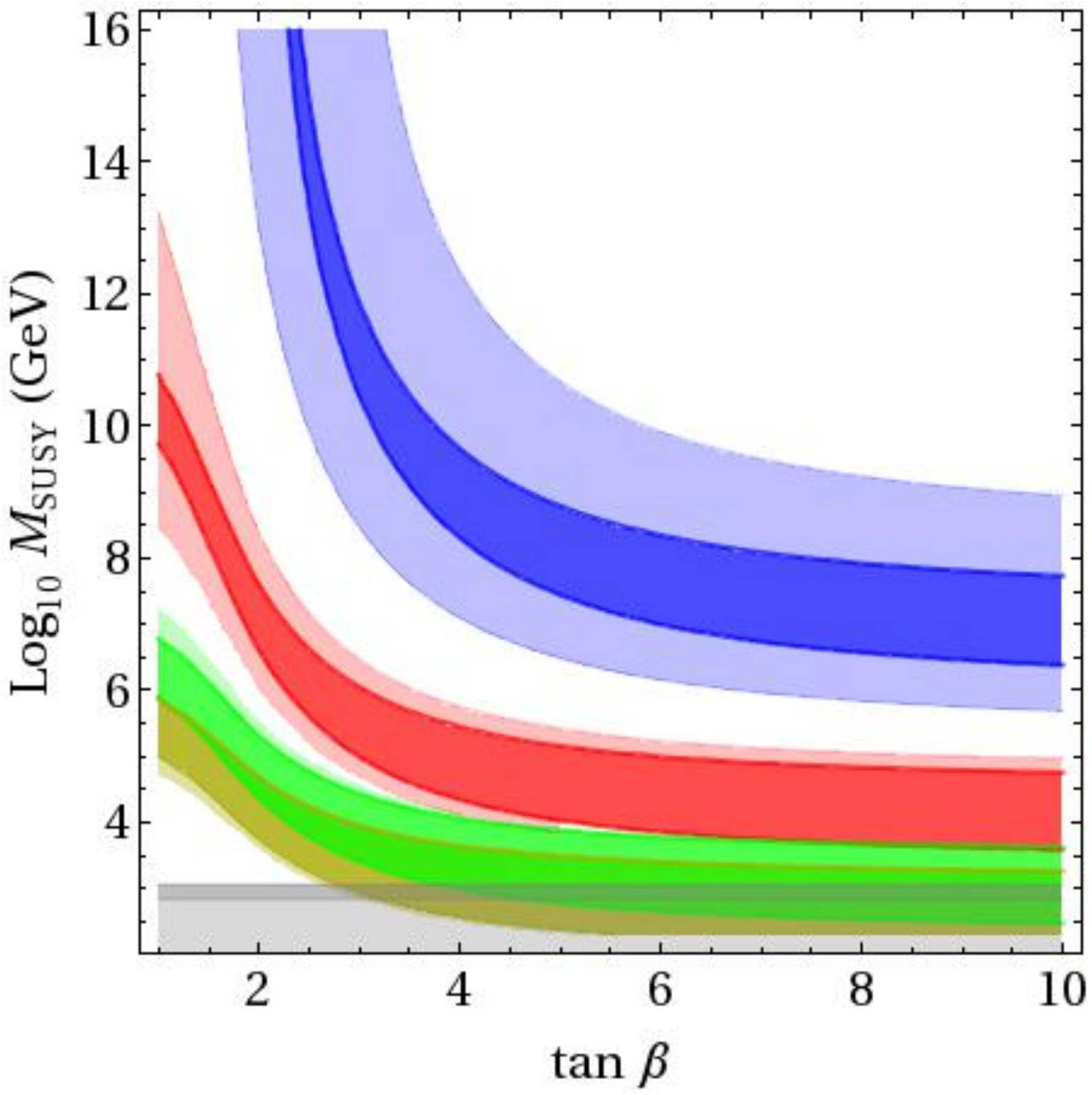}
\caption{Bands of constant $\tan\beta$ in the $M_{SUSY}-m_h$ (left panel) and constant Higgs mass in the $\tan\beta-M_{SUSY}$ plane (right panel) for the MSSM. From top to bottom $\tan\beta = 10\ ,3,\ 1$ and $m_h=140,130,120,115$ GeV respectively. The low (gray) horizontal band stands for the direct LHC lower bounds on $M_{\rm SUSY}$ (see text).}
\label{bound}
\end{center}
\end{figure}

Fig.~\ref{bound} (left panel) shows $m_h$ as a function of $M_{\rm SUSY}$ for three representative values of $\tan\beta$, namely $\tan\beta=1,3,10$. Due to the parametric dependence on $\cos^2 2\beta$ shown in eq.(\ref{mhaprox}) the results remain almost unchanged for larger values of $\tan\beta$. From the plots it becomes clear that one cannot reproduce an arbitrary large value of $m_h$ just by increasing $M_{\rm SUSY}$. Actually, there is an absolute upper bound of $\sim 145$ GeV, which becomes more stringent for moderate to low values of $\tan\beta$.
It should be kept in mind that $M_{\rm SUSY}$ essentially stands for ``stop masses". Indeed, in the usual MSSM scenarios the masses of all supersymmetric particles are of the same order (say, within a factor of 10 or less). Therefore, with this caveat, the following results are valid for essentially any MSSM model. A notable exception are split-SUSY models \cite{split}, where the masses of scalar superpartners are very high but the gauginos and higgsinos are still relatively light. Actually, the results for split-SUSY are analogous to those of the ordinary MSSM, but we will discuss them separately.

Now we can invert the previous argument and extract from the numerical results upper bounds on $M_{\rm SUSY}$ as a function of $\tan\beta$ and the upper (present and future) experimental bound on $m_h$.

These are given in Fig.~\ref{bound} (right panel), which shows the bands of constant Higgs mass in the $\tan\beta-M_{\rm SUSY}$ plane. The width of the bands comes from the various sources of uncertainty, to be discussed shortly. The behavior shown in Fig.1 can be {\em qualitatively} understood from  
the approximate expression (\ref{mass}). In particular, the larger $\tan\beta$ the bigger the first (tree-level) term in Eq.(~\ref{mass}) becomes, and thus a smaller value of $M_{\rm SUSY}$ is required to reproduce $m_h$. The width of each band (darker part) has been obtained by varying $X_t$ within its range, $0\leq X_t\leq 6$. Note that this uncertainty arises from our ignorance about the values of the remaining MSSM parameters. On top of that uncertainty, we have added the error coming from experimental uncertainties in the theoretical computation of $m_h$, resulting in the wider lighter bands. The experimental uncertainty is dominated by the one in the top mass $M_t=173.1\pm 1.25$ GeV. Additional sources of experimental error, such as the one in $\alpha_3$, are negligible when added in quadrature. 
Let us remark that we have {\em not} added in quadrature the uncertainties coming from the ignorance about $X_t$ and $M_t$, but linearly (to avoid statistical inconsistencies); thus the light band represents an overestimate of the total error. On the other hand, there is an intrinsic theoretical error coming from the higher-loop effects not considered in the computation of $m_h$, which we estimate in $\sim 2$ GeV \cite{Casas, Mariano, Haber}. This is negligible when added in quadrature to the other sources of error.

Now, each band of Fig.1 (right panel) represents the future upper bound on $M_{\rm SUSY}$, as soon as the upper bound on $m_h$ reaches the corresponding value. The present relevant bound, $m_h<149$ GeV, does not constrain the MSSM parameter space in a significant way. Actually, the corresponding band is outside Fig.1. But it is clear from the figure that as soon as new LHC bounds on $m_h$ are reported, the MSSM parameter space will become significantly cornered from above. Note also that the $m_h=115$ GeV band corresponds to the present {\em lower} bound on $M_{\rm SUSY}$. So in Fig.1 we see the future evolution of the MSSM window. On the other hand, if LHC discovers the Higgs, say at $m_h=130$ GeV, the associated band in Fig.1 gives the allowed region of the MSSM parameter space.

We have complemented the latter lower bound on $M_{\rm SUSY}$ with the direct lower bounds that LHC has already put on the MSSM parameter space \cite{LHCdirect}. They translate into the grey band at the bottom of Fig.1, which has been obtained as follows. LHC bounds on the MSSM are presented by ATLAS and CMS as exclusion regions in the constrained MSSM (CMSSM) parameter space,
The CMSSM is characterized by universality of the soft terms at $M_X$. 
We have extracted the data from the ATLAS exclusion plot with \texttt{EasyNData}~\cite{easyndata} and then calculated the exclusion contour for the gluino mass, $M_{\tilde g}$, versus $M_{\rm SUSY}$ with the \texttt{SoftSUSY} package \cite{softsusy}. It turns out that $M_{\rm SUSY}$ must be larger than 750-1000 GeV (the precise value depends on the value of $M_{\tilde g}$). This uncertainty in the bound is reflected in the narrow darker grey strip on top of the light one. Note that, strictly, the grey band corresponds to a lower bound for the CMSSM case; but one does not expects big changes in ordinary MSSM models. Certainly, in a general MSSM, stop masses (whose values determine $M_{\rm SUSY}$)) do not need to coincide with the squark masses of the first two generations (the ones directly probed so far by the LHC). However, this does not mean that stop masses can be anything. Even taking a vanishing initial stop mass at the scale where SUSY breaking is communicated to the observable sector, say $M_X$, radiative corrections produce a contribution $\sim 2.5\,M_{\tilde g}^2$. If the breaking of SUSY is communicated at a lower scale this contribution decreases logarithmically with the scale but it is always very substantial. This does not mean that one of the stops might not be significantly lighter, due to the potential large mixing between them, but the average, i.e. $M_{\rm SUSY}$, is always large. Consequently the LHC direct lower bounds on $M_{\rm SUSY}$, represented by the grey band in Fig.~\ref{bound}, remains valid for most of the MSSM models (the only exception would be models with large splitting of the third generation and very low scale of SUSY breaking communication).

One could take the attitude of only considering low $M_{\rm SUSY}$, say $M_{\rm SUSY}\lsim {\cal O}({\rm TeV})$, as reasonable, in order to avoid fine-tuning to get the correct electroweak scale. Then, some of the upper bounds shown in Fig.~\ref{bound} would be irrelevant. However, it has been suggested that in a landscape scenario such fine-tuning can be largely compensated by the overabundance of vacua with SUSY broken at a high scale, in which the anthropic principle would operate, see e.g. ref. \cite{susskind}. As we have seen, this kind of scenario is going to be tested by LHC very soon. The split-SUSY framework \cite{split}, which we are going to discuss next, is in fact a popular variant of the above-mentioned landscape scenario.

\begin{figure}[t]
\begin{center}
\includegraphics[width=0.45\linewidth]{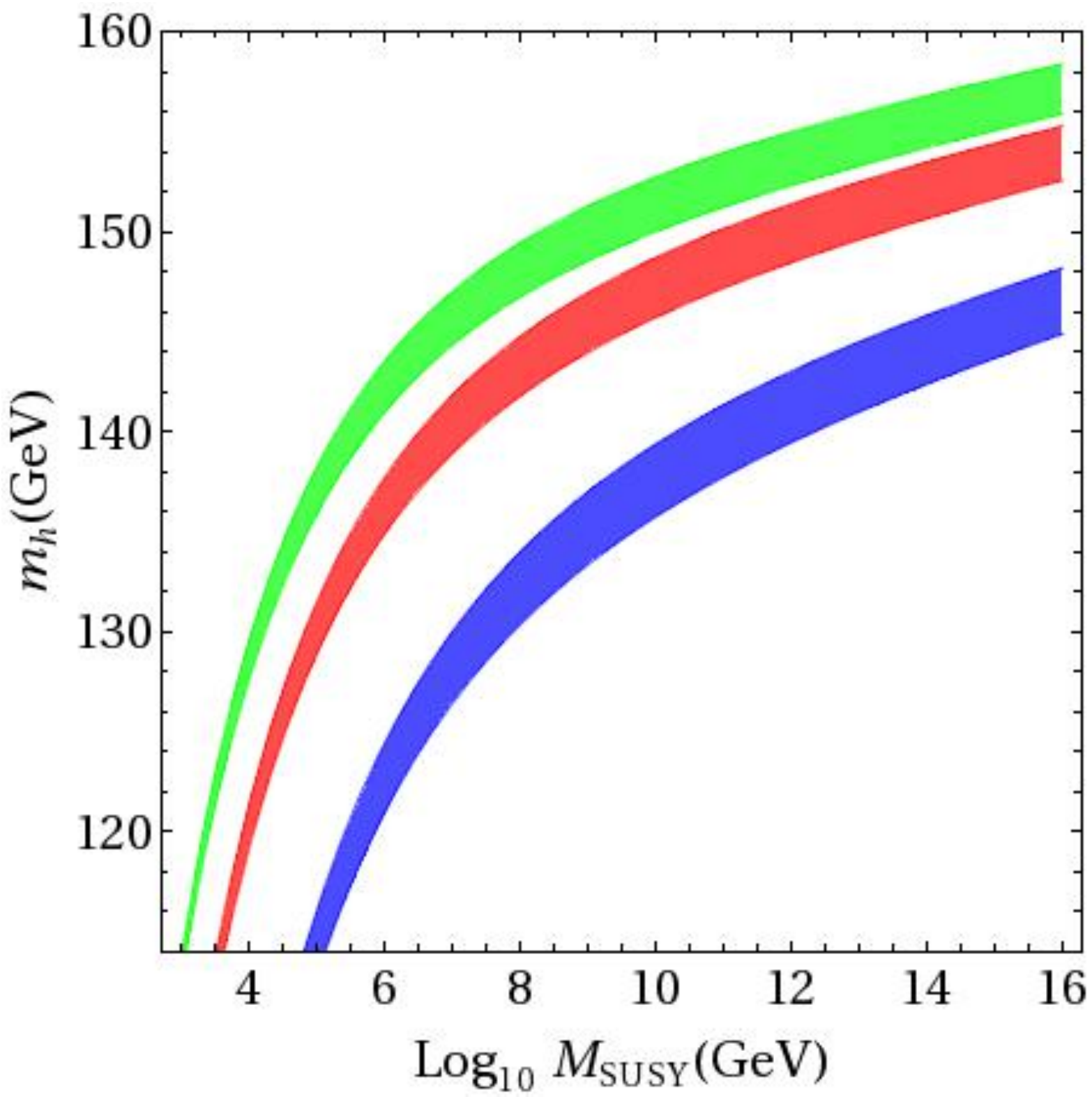}%
\hspace{0.05\linewidth} %
\includegraphics[width=0.45\linewidth]{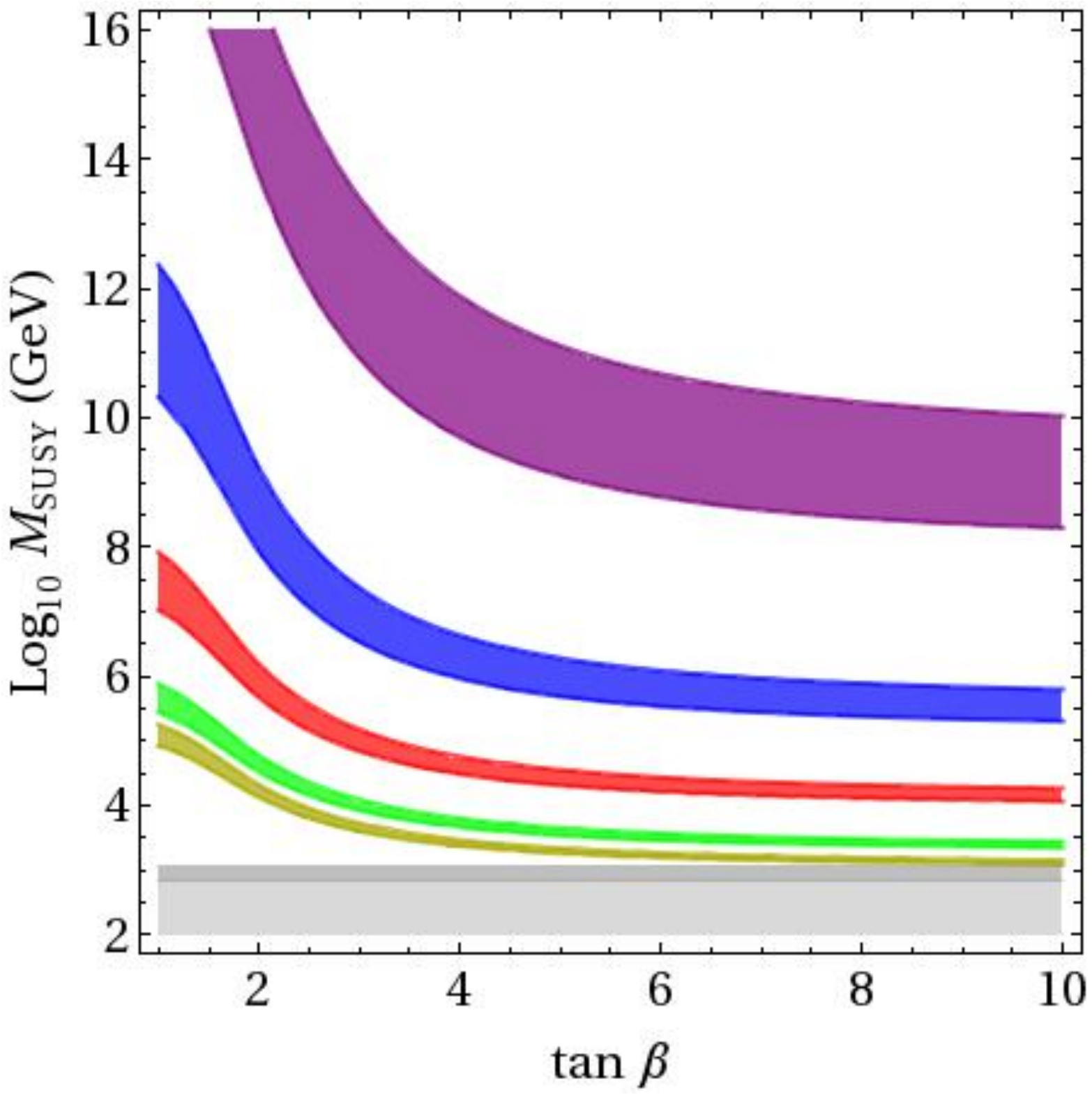}
\caption{The same as Fig.1 but for split-SUSY. In the $\tan\beta-M_{SUSY}$ plane, from top to bottom $m_h=150, 140, 130, 120, 115$ GeV. }
\label{bound-split}
\end{center}
\end{figure}

In split-SUSY one supposes that the spectrum of superpartners is split, having scalars at higher scales whereas gauginos and higgsinos remain light, usually thanks to an approximate R-symmetry. In this way, split-SUSY is kept consistent with the perturbative unification of gauge couplings at $M_X$ and still contains Dark Matter candidates. The approximate R-symmetry is also responsible to keep $A-$ and $\mu-$parameters small, something necessary for the radiative stability of the scenario. So in split-SUSY there are two well-separated SUSY thresholds (thus the name).The prediction for $m_h$ in split-SUSY is done following a similar approach to the one previously discussed and applied for the MSSM. This prediction was already considered in the second seminal split-SUSY paper of ref. \cite{split}. We have updated that calculation (using e.g. upgraded information about the top mass) and studied in  a systematic way the bounds on $M_{\rm SUSY}$ depending the evolution of the upper bound on $m_h$. The main difference with respect to the MSSM case is that in split-SUSY the gluinos remain active and contribute significantly to the RG equations between the upper and the lower SUSY-thresholds.

The results are shown in Fig.~\ref{bound-split}, which is analogous to the previous Fig. 1 for the MSSM. Due to the smallness of $A_t$ and $\mu$ (and thus of $X_t$), the threshold correction for $\lambda$ at the high SUSY threshold is negligible, see Eq.~(\ref{lambda}).
In consequence, the width of the bands in Fig. 2 is only due to the experimental error on $M_t$. Note also that $M_{\rm SUSY}$ corresponds to the {\rm upper} SUSY-threshold ($\sim$ stop masses), and therefore the plotted upper bounds are absolute upper bounds for SUSY in the split-SUSY scenario. In contrast to the MSSM case, there is a region of the parameter space, large $\tan\beta$ and very heavy masses, that is already excluded with today's bound on the Higgs mass; yielding $M_{\rm SUSY}<10^{11}$ GeV. As soon as the upper bound on $m_h$ reaches 140 GeV, the exclusion will hold for any $\tan \beta$. This is most relevant for split-SUSY, since in this scenario we {\em do expect} the upper SUSY-threshold to lie at very high energy; typically as large as $M_X$ or $M_p$, something which is going to be probed very soon by LHC.

To conclude, we have used the current and forthcoming LHC upper bounds on the Higgs mass to put upper bounds on supersymmetric masses, $M_{\rm SUSY}$; using the fact that in the MSSM the quartic Higgs coupling, and therefore the Higgs mass, is a function of the SUSY masses (in particular stop masses). Right now there is no significant constraint on the parameter space of the MSSM but very soon there will be, as can be seen in Fig.~\ref{bound}. On the other hand, for split-SUSY a non-negligible part of the parameter space can already be excluded on these grounds, as can be seen in Fig.~\ref{bound-split}. As the LHC produces better bounds on $m_h$ the allowed region will be more and more shrunk, therefore showing that, even if SUSY is not found, it can not be hidden way. 
Eventually (and hopefully) the Higgs will be discovered, and one can use these results to establish the region of the $M_{\rm SUSY}-\tan\beta$ plane consistent with the actual value of $m_h$.

\section*{Acknowledgments}
 This work was partly supported by the National Science Foundation under grant PHY-0905383-ARRA; and by the MICINN, Spain, under contracts FPA2010-17747 and Consolider-Ingenio PAU CSD2007-00060. We thank as well the Comunidad de Madrid through Proyecto HEPHACOS ESP-1473. M. E. Cabrera acknowledges the financial support of the CSIC through a predoctoral research grant (JAEPre 07 00020).

\bibliographystyle{unsrt}

\end{document}